%% file: direct.tex
\documentclass[a4paper,10pt]{article}
\usepackage{amsmath,amssymb,epsfig,graphicx} 
\usepackage{setspace}

\begin{document}

%\onehalfspacing
%\doublespacing

\title{Consistent simulation of direct-photon production in hadron collisions 
including associated two-jet production}

\author{Shigeru Odaka and Yoshimasa Kurihara\\
High Energy Accelerator Research Organization (KEK)\\
1-1 Oho, Tsukuba, Ibaraki 305-0801, Japan\\
E-mail: \texttt{shigeru.odaka@kek.jp}}

\date{}

\maketitle

%--------1---------2---------3---------4---------5---------6---------7---------8
\begin{abstract}
We have developed an event generator for direct-photon production 
in hadron collisions, including associated two-jet production 
in the framework of the GR@PPA event generator.
The event generator consistently combines $\gamma$ + 2-jet production 
processes with the lowest-order $\gamma$ + jet and
photon-radiation (fragmentation) processes from QCD 2-jet production
using a subtraction method.
The generated events can be fed to general-purpose event generators 
to facilitate the addition of hadronization and decay simulations.
Using the obtained event information, 
we can simulate photon isolation and hadron-jet reconstruction 
at the particle (hadron) level.
The simulation reasonably reproduces measurement data obtained at the LHC 
concerning not only the inclusive photon spectrum, but also the correlation 
between the photon and jet. 
The simulation implies that the contribution of the $\gamma$ + 2-jet is 
very large, especially in low photon-$p_{T}$ ($\lesssim$ 50 GeV) regions.
Discrepancies observed at low $p_{T}$, although marginal, may indicate 
the necessity for the consideration of further higher-order processes.
Unambiguous particle-level definition of the photon-isolation condition 
for the signal events is desired to be given explicitly in future measurements. 
\end{abstract}

%% main text
\section{Introduction}
\label{sec:intro}

High-energy isolated photon production in hadron collisions 
is considered to be suitable for probing 
the dynamics of partons (quarks and gluons) 
inside hadrons~\cite{d'Enterria:2012yj,Carminati:2012mm}, 
as such photons are expected to be produced via parton-level 
quantum-electrodynamic (QED) interactions, 
referred to as direct-photon production.
In addition, an understanding of this process is important for 
hadron-collision experiments because various new phenomena can be explored 
by observing events involving energetic isolated photon(s).
In fact, photon detection played a crucial role in the discovery 
of the Higgs boson~\cite{Aad:2012tfa,Chatrchyan:2012ufa}.

\begin{figure}[tp]
\begin{center}
\includegraphics[scale=0.5]{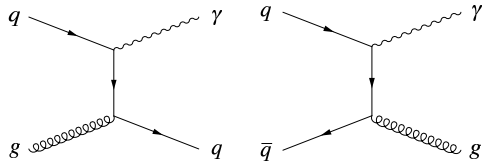}
\caption{\label{fig:diagram1}
Lowest-order diagrams of direct-photon production in hadron collisions.
}
\end{center}
\end{figure}

\begin{figure}[tp]
\begin{center}
\includegraphics[scale=0.5]{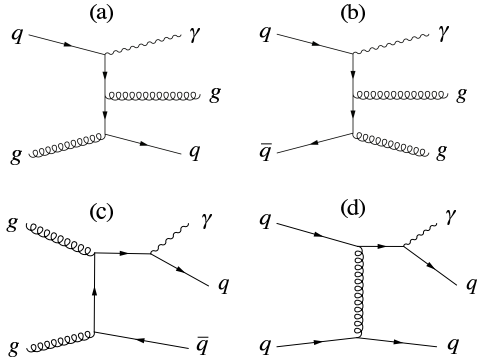}
\caption{\label{fig:diagram2}
Typical Feynman diagrams of $\gamma$ + 2-jet production. 
The diagrams (a) and (b) are simple gluon-radiation corrections 
to the diagrams in Fig.~\ref{fig:diagram1}, 
while (c) and (d) are new processes that first emerge at this order.
}
\end{center}
\end{figure}

The lowest-order processes of direct-photon production are the $\gamma$ + jet 
production processes illustrated in Fig.~\ref{fig:diagram1}, 
in which the produced high-energy quark or gluon is likely to resolve 
to a hadron jet.
However, these processes cannot sufficiently explain 
reported measurement results.
A next-to-leading-order (NLO) calculation by JETPHOX~\cite{Catani:2002ny} 
has shown that the next-order $\gamma$ + 2-jet contribution is very large;
the correction to the lowest order amounts to more than 100\% 
under typical measurement conditions.
Typical diagrams of the $\gamma$ + 2-jet production processes are shown 
in Fig.~\ref{fig:diagram2}.
Together with the simple correction diagrams, (a) and (b),
new processes, (c) and (d), appear at the same coupling order.
These new processes may make large contributions, 
as a result of the high gluon density in hadrons for the process (c)　
and the variety of quark combinations for (d).
Owing to the large contribution of the $\gamma$ + 2-jet, 
NLO predictions still have relatively large energy-scale dependences.

The CERN LHC has already provided high-statistics direct-photon data 
from proton-proton collisions to the 
ATLAS~\cite{Aad:2010sp,Aad:2011tw,ATLAS:2012ar,Aad:2013gaa,Aad:2013zba} 
and CMS~\cite{Khachatryan:2010fm,Chatrchyan:2011ue,Chatrchyan:2013mwa} 
experiments.
Although the statistics is very high, 
the measurement precision is still limited.
This is because the measurements suffer from the effects of a large 
hadron-decay background. 
As high-energy hadrons are produced in a jet-like structure, 
this background can be dramatically reduced by imposing an isolation condition 
on the photon candidates.
The isolation condition is usually defined using the sum of the transverse 
energies of particles inside a cone around the photon ($E_{T}^{\rm cone}$).
The cone size ($\Delta R$) is usually defined by the quadratic sum of 
the separations in the azimuthal angle ($\phi$) and pseudorapidity ($\eta$) 
as $\Delta R^{2} = \Delta\phi^{2} + \Delta\eta^{2}$, 
and $E_{T}^{\rm cone}$ is required to be smaller than a certain value as
\begin{equation}\label{eq:iso}
  E_{T}^{\rm cone} < E_{T}^{\rm iso} . 
\end{equation}
Although the reduction obtained using this approach is very effective, 
relatively large systematic uncertainties remain 
because the remaining background events that mimic the direct-photon production 
are very unusual and less known.
Together with the remaining energy-scale dependence in the NLO calculations, 
these uncertainties currently limit the precision of the comparison with 
theoretical predictions to the 10\% level.
Despite imposed large uncertainties, however, agreement between prediction 
and experimental results is marginal in the measurements reported to date, 
although obvious discrepancies have not been found. 

The measurement uncertainties will be reduced as our understanding 
of the background is improved, 
and theoretical predictions will become more accurate through the inclusion 
of further higher orders.
However, in anticipation of comparisons with accuracy better than 10\%, 
we have some concerns regarding the signal definition 
which is relevant to both the background and efficiency estimations.
As we can see in the diagrams shown in Fig.~\ref{fig:diagram2}, 
the $\gamma$ + 2-jet processes have final-state QED collinear divergences.
The isolation conditions imposed to reduce the background also strongly 
reject $\gamma$ + 2-jet events 
in which the photon is produced almost in parallel to a quark.
Those events to be rejected are not interesting in terms of 
the direct-photon measurement.
In addition, inclusion of the isolation condition can reduce the uncertainties 
concerning non-perturbative effects which should regularize the collinear 
divergence.
Hence, it is preferable to include the isolation condition in the definition 
of signal events for which measurement results and theoretical predictions 
are presented.

In order to conduct rigorous comparisons between measurements and predictions, 
we must define the isolation condition unambiguously.
The detection efficiency must be estimated for events satisfying 
the isolation condition, 
and those events that do not satisfy the isolation condition must be treated 
as background, even if they are produced via one of the processes 
shown in Fig.~\ref{fig:diagram2}.
Usually, 
theoretical calculations apply isolation conditions at the parton level. 
However, in principle, it is impossible to give unambiguous definitions 
at the parton level since partons are not physical objects.
The appearance of partons depends on the perturbation order 
and arbitrary cutoffs.
Unambiguous definitions can be given only by using particle(hadron)-level 
information, as in the hadron-jet definition~\cite{Buttar:2008jx}.
Such a definition has already been adopted 
in recent measurements~\cite{Aad:2011tw,Aad:2013zba}, 
in which the particle species to be included for the evaluation of 
$E_{T}^{\rm cone}$ are also explicitly specified.

If the definition is given at the particle level, 
we must therefore convert parton-level theoretical predictions into 
predictions at the particle level.
Although it may be possible to evaluate relevant corrections 
using separate simulations, 
the most straightforward approach is to use parton-level Monte Carlo (MC) 
event generators that can be connected to simulations 
down to the particle level. 
In such simulations, it is preferable that the event generation is based on 
matrix elements (MEs) consistently including multi-jet production processes, 
in order to achieve sufficient precision.

We have been developing event generators that facilitate the matching 
between processes having different jet-multiplicities, 
based on a subtraction method~\cite{Kurihara:2002ne,Odaka:2007gu} in the 
framework of the GR@PPA event generator~\cite{Tsuno:2002ce,Tsuno:2006cu}.
The generated events can be fed to general-purpose event generators, 
such as PYTHIA~\cite{Sjostrand:2006za} and HERWIG~\cite{Corcella:2000bw}, 
in order to simulate hadronization and decays 
to obtain particle-level event information.
We first developed event generators that combine weak-boson production 
processes associated with 0-jet and 1-jet 
production~\cite{Odaka:2011hc,Odaka:2012da}, 
and demonstrated that the developed event generator reproduces measured 
$Z$-boson $p_{T}$ spectra very 
precisely~\cite{Odaka:2009qf,Odaka:2012iz,Odaka:2013fb}.
In these event generators, 
collinear divergent components of quantum-chromodynamics (QCD) are subtracted 
from 1-jet MEs in order to prevent double-counting with the jet production 
in parton-shower (PS) simulations applied to 0-jet events. 
The matching method has been extended to diphoton ($\gamma\gamma$) 
production~\cite{Odaka:2012ry}, in which the subtraction is applied 
to both QED and QCD collinear divergences.
Recently, we have succeeded in combining 2-jet production processes 
with 0- and 1-jet production~\cite{Odaka:2014ura}, 
where the subtraction method is modified to account for 
the soft-gluon divergence together with the collinear divergence.

In this article, 
we construct an event generator that consistently combines the simulation of 
$\gamma$ + 2-jet production with the lowest-order $\gamma$ + 1-jet 
using the techniques that we have developed to date.
QCD 2-jet production having hard-photon radiation from PS is also incorporated 
in order to restore the subtracted QED collinear components.
The generated events, to which further simulations down to the particle level 
are added, are compared with measurement results at the LHC.
Although the event generation is based on tree-level MEs only, 
we can expect the simulations to have precision comparable to NLO calculations, 
as the correction to the lowest order is expected to be dominated by 
$\gamma$ + 2-jet real radiation contributions.
Subsequently, we present other distributions that can be derived from 
the simulation results, in order to discuss the characteristic properties 
of direct-photon production.
The understanding of this process obtained through such studies will be 
helpful in reducing the systematic uncertainties in future measurements.

Note that a similar event generator has already been developed 
in the SHERPA framework~\cite{Hoeche:2009xc}, and 
its simulation results reproduce the event kinematics of diphoton 
($\gamma\gamma$)~\cite{Aad:2012tba} and direct-photon~\cite{Aad:2013gaa} 
production with good precision.
In the SHERPA event generator, processes having different jet-multiplicities 
are merged using the CKKW method~\cite{Catani:2001cc}, 
in which multi-jet events generated according to jet-including MEs are 
suppressed by reinterpreting them in the context of a PS.
The suppression is applied without separating non-divergent components from 
divergent components that PS is based on.
Such an ME-based simulation is applied down to a small merging scale, 
and application of PS is limited to further soft regions.
We are taking a different approach to achieving the same goal by using 
the subtraction method.
In our approach, the subtraction and PS simulations are applied 
up to the typical energy scale of events for which MEs are evaluated.
Thus, no suppression is required and non-divergent components in MEs 
are preserved as they are in the entire phase space.

The rest of this article is organized as follows.
The overall simulation strategy is discussed in Sec.~\ref{sec:strategy}, 
and the self-consistency of the developed event generator is examined 
in Sec.~\ref{sec:matching}.
We compare the simulation results with LHC measurements 
in Sec.~\ref{sec:comparison}.
Using the simulated events, the characteristic properties of direct-photon 
production are discussed in Sec.~\ref{sec:discussion}, 
and the discussions are concluded in Sec.~\ref{sec:conclusion}.

\section{Simulation strategy}
\label{sec:strategy}

Programs for calculating the MEs of the $\gamma$ + 2-jet production processes 
were newly generated using the GRACE system~\cite{Ishikawa:1993qr,Yuasa:1999rg}
and installed in the GR@PPA event generator framework.
The MEs are based on Feynman diagrams including all possible combinations of 
quarks (up to the $b$ quark) and gluons having photon radiation from one of 
the quark lines.
Thus, the QCD part of the process is identical to the full set of 
QCD 2-jet production processes from which $gg \rightarrow gg$ is excluded.

The $\gamma$ + 2-jet MEs have various collinear divergences.
We subtract these divergences using the limited leading-logarithmic (LLL) 
subtraction method~\cite{Odaka:2011hc} to render the cross sections finite.
Together with the QCD divergences,
final-state QED divergences are subtracted using a method 
that has been developed previously~\cite{Odaka:2012ry}.
We do not need to consider initial-state QED divergences 
since an energetic photon is required to be detected at large angles.
In the subtraction of QCD divergences, we also need to account 
for the soft-gluon divergence in the $\gamma$ + 2-jet production.
This divergence is subtracted simultaneously with the collinear divergences, 
using the combined subtraction method 
developed in our previous study~\cite{Odaka:2014ura}.

The subtraction is applied in order to prevent double-counting of 
the jet production in MEs and that from PSs attached to 
lower jet-multiplicity events.
Thus, the subtracted components are restored by combining 
lower multiplicity processes to which PS simulations are applied. 
The subtracted QCD components are restored by combining $\gamma$ + 1-jet 
production processes.
In accordance with the modification to the subtraction designed to 
account for the soft-gluon divergence, 
the $\gamma$ + 1-jet events are reweighted 
after PS simulations are applied~\cite{Odaka:2014ura}.

The subtracted QED components are restored by combining QCD 2-jet production 
processes to which a QCD/QED-mixed PS is applied~\cite{Odaka:2012ry}. 
The photon radiation in low-$Q^{2}$ regions, 
which PS simulations do not accommodate, 
is simulated by using a fragmentation function (FF)~\cite{Bourhis:1997yu} 
as in the diphoton-production simulation~\cite{Odaka:2012ry}.
In order to improve the generation efficiency, 
the production of one energetic photon is enforced in this PS/FF simulation.
The enforcing procedure determines an additional event weight for 
the simulated events.

The event generation is performed using the LabCut framework 
of GR@PPA~\cite{Odaka:2011hc}.
In this framework, 
PS simulations are applied before differential cross section values are 
passed to the integration and event generation utility, 
BASES/SPRING~\cite{Kawabata:1985yt,Kawabata:1995th}, employed in GR@PPA.
Thus, the cross section values can be weighted using event weights 
that are determined after application of the PS simulations.
In the generation of $\gamma$ + 1-jet and QCD 2-jet events, 
the evaluated event weights are accounted for in the integration and 
event generation using this framework.

The LabCut framework was originally introduced for the efficient generation 
of events for which strict kinematical constraints are required. 
As PS simulations alter the momenta of generated particles, 
the constraints in parton-level hard interactions must be relaxed 
significantly in order to avoid biases to the final sample.
Direct-photon production is such a process.
For instance, when we generate the events used in the next section, 
in which a $p_{T}$ cut of 100 GeV/$c$ is applied to the photon, 
we must reduce the threshold to 40 GeV/$c$ in the hard-interaction generation.
We would suffer from extremely poor efficiency in the event selection 
if we adopted an ordinary generation scheme.
For this reason, we also adopt the LabCut framework for the generation 
of $\gamma$ + 2-jet events.

Since PS simulations are limited by a certain $Q^{2}$ value 
($Q^{2} < \mu_{\rm PS}^{2}$), 
the subtraction must also be limited by the same $Q^{2}$ 
in order to achieve matching between them.
However, setting the limit for the LLL subtraction ($\mu_{\rm LLL}$) equal to 
$\mu_{\rm PS}$ is not trivial, 
since these values are determined in different processes.
In order to solve this problem, GR@PPA includes a mechanism that refers to 
energy scales defined for different processes.
In the LLL subtraction, 
$\mu_{\rm PS}$ of the non-radiative event is referred to 
using event information that is reconstructed by removing the assumed radiation, 
in order to assign the same value to $\mu_{\rm LLL}$.
Thus, in principle, 
any energy-scale definition can be adopted to achieve matching.

The above energy scale may differ for the initial-state radiation (ISR) 
and the final-state radiation (FSR).
Hence, we denote these boundaries as $\mu_{\rm ISR}$ and $\mu_{\rm FSR}$, 
respectively.
In the event generation, 
we must also specify the other energy scales necessary for QCD calculations, 
{\it i.e.}, the renormalization scale ($\mu_{R}$) 
and the factorization scale ($\mu_{F}$).
As we require a basic identity between the initial-state PS and 
the QCD evolution in parton distribution functions (PDFs), 
$\mu_{\rm ISR}$ must be equal to $\mu_{F}$ in our matching method.
In principle, $\mu_{R}$ may differ from $\mu_{F}$.
However, since $\alpha_{s}$ running is assumed in PS and PDF, 
it would be better to choose an identical value in order to achieve 
an equal jet-production strength at the boundary between PS and ME.
For these reasons, 
we use an identical value for $\mu_{R}$, $\mu_{F}$, and $\mu_{\rm ISR}$ 
as the default setting.
Although there is no strict reason, 
we also adopt the same value for $\mu_{\rm FSR}$ in the present study, 
for simplicity.
Thus, the energy scale is unique in the present study, {\it i.e.},
$\mu = \mu_{R} = \mu_{F} = \mu_{\rm ISR} = \mu_{\rm FSR}$.

We must then decide how to define $\mu$.
As we have discussed above, 
any definition is acceptable for the matching.
However, a uniform definition that is applicable to all considered processes 
is preferable.
Direct-photon production is considered to be dominated by $t$-channel 
interactions.
Hence, it would be natural to use the $p_{T}$ value of the photon 
for $\mu$.
However, the photon is not necessarily the hardest particle 
in $\gamma$ + 2-jet production.
Moreover, no photon appears in QCD 2-jet production. 

In the present study, 
we adopt a new definition that can be applied to all considered processes, 
\begin{equation}\label{eq:scale}
  \mu = {\rm max}\{ Q_{T,i} \} .
\end{equation}
The quantity $Q_{T,i}$ is defined for each final-state particle 
in terms of its mass and $p_{T}$, such that
\begin{equation}\label{eq:qt}
  Q_{T,i}^{2} = m_{i}^{2} + p_{T,i}^{2} , 
\end{equation}
and the largest value is taken as the energy scale of the event, 
as defined in Eq.~(\ref{eq:scale}).
This definition reduces to $\mu = m$ for single-resonance production 
and to $\mu^{2} = m^{2} + p_{T}^{2}$ for pair production.
It also reduces to $\mu^{2} = m_{V}^{2} + p_{T}^{2}$ 
for vector boson ($V$) + jet production processes.
In the processes considered in the present study, 
$\mu$ is determined by the $p_{T}$ value of the particle 
having the largest $p_{T}$ since all particles are nearly massless.

The generated event samples of $\gamma$ + 2-jet ($a2j$), 
$\gamma$ + 1-jet ($a1j$), and QCD 2-jet ($qcd2j$),
to which PS simulations are fully applied down to $Q_{0}$ = 5 GeV, 
are passed to PYTHIA 6.425~\cite{Sjostrand:2006za} 
in order to add hadronization and decay simulations.
Although low-$Q^{2}$ ($< Q_{0}^{2}$) QCD effects are also simulated in PYTHIA, 
their effects are almost negligible in terms of the observable quantities 
of direct-photon events.
However, the momenta of the generated particles are more or less altered 
by these simulations.
In order to avoid possible biases, 
the generation conditions for the events to be passed to PYTHIA 
are slightly relaxed from the final selection criteria.
For instance, accounting for a large safety margin, 
the $p_{T}$ threshold of photons is reduced to 90 GeV/$c$ 
for the final selection condition of $p_{T} >$ 100 GeV/$c$.
The PYTHIA simulation is applied with its default setting, 
except for the settings of {\tt PARP(67) = 1.0} and {\tt PARP(71) = 1.0}
as in the previous studies~\cite{Odaka:2009qf,Odaka:2012iz,Odaka:2013fb}.
In addition, 
the underlying-event simulation is deactivated ({\tt MSTP(81) = 0}) 
since LHC experiments correct the isolation cone energy for this effect.

As described earlier, 
the PS simulations in GR@PPA rotate and boost the original events 
generated by hard interactions significantly.
In order to check possible biases due to the hard-interaction generation 
conditions, the event information prior to application of the PS simulations 
is passed to PYTHIA, together with the PS-applied event information. 
The spectra relevant to the original generation conditions are checked 
after applying the particle-level event selection, 
in order to confirm that the selected sample is not biased by 
the generation conditions.

\section{Matching test}
\label{sec:matching}

The self-consistency of the developed event generator was tested 
by investigating observable quantities in the simulated events.
The events were generated under the 7-TeV LHC condition, {\it i.e.}, 
for proton-proton collisions at a center-of-mass (CM) energy of 7 TeV,
using the CTEQ6L1 PDF~\cite{Pumplin:2002vw} included in the GR@PPA package.
We required that the events simulated down to the particle level 
contain the photon generated by GR@PPA in the transverse-momentum ($p_{T}$) 
and pseudorapidity ($\eta$) ranges of 
\begin{equation}\label{eq:photon1}
  p_{T}(\gamma) > 100 {\rm ~GeV}/c {\rm ~~and~~} |\eta(\gamma)| < 2.4 .
\end{equation}
As particle-level event information is available, 
we can apply a realistic isolation condition to the photons.
We adopted the simplest definition in Eq.~(\ref{eq:iso}).
The cone $E_{T}$ ($E_{T}^{\rm cone}$) was evaluated 
using all stable particles, excluding neutrinos and muons, 
produced inside the cone around the photon having a size of $\Delta R$ = 0.4.
We defined the isolation condition as
\begin{equation}\label{eq:iso1}
  E_{T}^{\rm cone} < 7.0 {\rm ~GeV} .
\end{equation}

We need hadron-jet information in the simulated events in order to test 
the matching of the $\gamma$ + 2-jet production with other processes.
The hadron jets were reconstructed using FastJet 3.0.3~\cite{Cacciari:2011ma}, 
with the application of the anti-$k_{T}$ algorithm with $R = 0.4$.
All stable particles in the pseudorapidity region of $|\eta| < 5.0$, 
including neutrinos, were used for the reconstruction.
The reconstructed jets that satisfied the conditions,
\begin{equation}\label{eq:jet1}
  p_{T}({\rm jet}) > 30 {\rm ~GeV}/c ,  ~~|\eta({\rm jet})| < 4.4 ,  
  {\rm ~~and~~}\Delta R(\gamma, {\rm jet}) > 0.5 ,
\end{equation}
were taken as the detected jets to be used in the subsequent analyses, 
where $\Delta R(\gamma,{\rm jet})$ is the separation in $\Delta R$, 
$\Delta R^{2} = \Delta\phi^{2} + \Delta\eta^{2}$, between the photon and jet.
We required that at least two jets were detected in each event.

Since $\mu$ defines the boundaries between the PS and 
ME-based simulations for divergent components, 
a strict test of the matching can be conducted by investigating 
the stability of observable distributions against the variation of $\mu$.
We take the scale defined by Eq.~(\ref{eq:scale}) 
as the standard value $\mu_{0}$.
The simulations were carried out for extreme choices of $\mu = \mu_{0}/2$ 
and $\mu = 2\mu_{0}$, together with the standard setting of $\mu = \mu_{0}$.
The $p_{T}$ threshold of photons in Eq.~(\ref{eq:photon1}) was selected 
in order to ensure that $\mu$ is always larger than the $p_{T}$ threshold 
of hadron jets in Eq.~(\ref{eq:jet1}).
Hence, events are always allowed to include {\it soft} hadron jets 
that are predominantly produced by PS simulations.

Because of a large contribution from the $\gamma$ + 2-jet, 
the integrated cross section has a significant $\mu$ dependence, 
yielding a 37\% increase for $\mu = \mu_{0}/2$ 
and an 18\% decrease for $\mu = 2\mu_{0}$ 
with respect to the value obtained for the standard setting.
Nearly two-thirds of the variation is caused by the change in $\mu_{R}$ 
and the remainder is due to $\mu_{F}$.
As these changes are not relevant to the matching to be tested here, 
we compare the simulation results in terms of the normalized differential 
cross section, $(1/\sigma)d\sigma/dx$, where $x$ denotes the tested quantity.

\begin{figure}[tp]
\begin{center}
\includegraphics[scale=0.6]{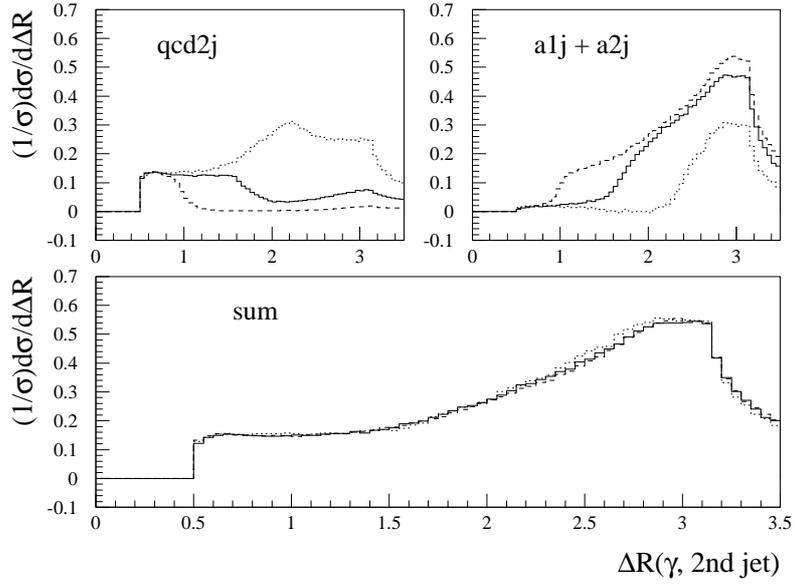}
\caption{\label{fig:matching1}
Distribution of the $\Delta R$ separation between the photon and second jet.
The distributions of $qcd2j$ and $a1j + a2j$ are shown separately 
in the upper panels, and the sum is shown in the lower panel.
The results for $\mu = \mu_{0}/2$, $\mu_{0}$, and $2\mu_{0}$ are 
represented by dashed, solid, and dotted histograms, respectively.
}
\end{center}
\end{figure}

The matching for the final-state photon radiation can be tested 
by investigating the correlation between the photon and the second jet, 
{\it i.e.}, the detected hadron jet having the second-largest $p_{T}$.
The distribution of the $\Delta R$ separation between them is shown 
in Fig.~\ref{fig:matching1}.
The $qcd2j$ result and the sum of the $a1j$ and $a2j$ results, 
labeled $a1j + a2j$, are shown separately in the upper panels.
The photon is produced by PS in the former case, 
while it is produced in the hard interaction in the latter.
Events in the $\Delta R < 0.5$ region are rejected by the requirement 
imposed on the jets in Eq.~(\ref{eq:jet1}).
The $\mu = \mu_{0}$ results are illustrated using solid histograms, 
and those for $\mu = \mu_{0}/2$ and $\mu = 2\mu_{0}$ are represented by 
dashed and dotted histograms, respectively.
We can see that these two results have remarkably large $\mu$ dependences.
The sum of the results for all simulated processes is shown in the lower panel.
The large $\mu$ dependences observed in the separate results are compensated 
by each other to form a very stable distribution in the summed result, 
implying that almost complete matching is accomplished.

\begin{figure}[tp]
\begin{center}
\includegraphics[scale=0.6]{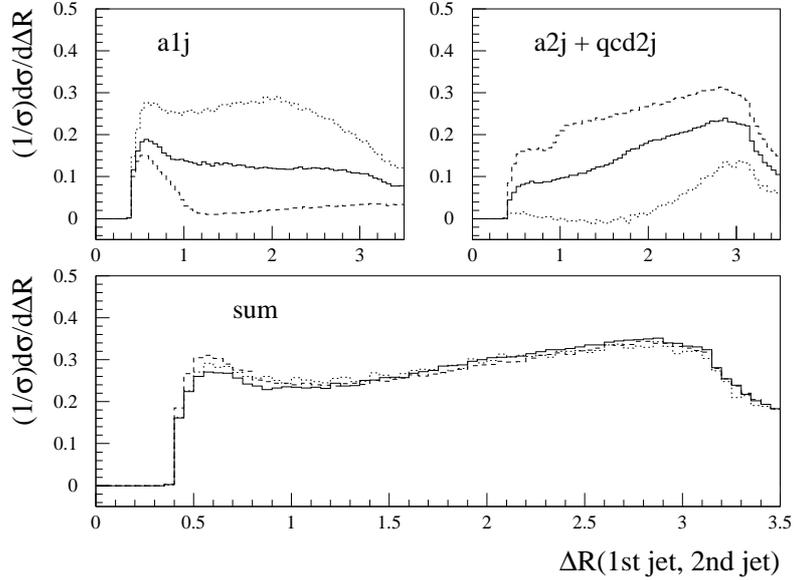}
\caption{\label{fig:matching2}
Distribution of the $\Delta R$ separation between the leading and second jets.
The distributions for $a1j$ and $a2j + qcd2j$ are shown separately 
in the upper panels, and the sum is shown in the lower panel.
The results for $\mu = \mu_{0}/2$, $\mu_{0}$, and $2\mu_{0}$ are 
represented by dashed, solid, and dotted histograms, respectively.
}
\end{center}
\end{figure}

Analogously, $\Delta R$ separation between the leading (highest-$p_{T}$) 
and second jets is a good measure to test the matching 
in the final-state QCD radiation.
This $\Delta R$ distribution is shown separately for $a1j$ and $a2j + qcd2j$ 
in the upper panels in Fig.~\ref{fig:matching2}.
The second jet is produced by PS in the former, while the hard interaction 
contains two jets in the latter.
The sum of these results is shown in the lower panel.
The events are not distributed in the $\Delta R < 0.4$ region 
because of the cone size in the jet reconstruction.
Again, although the dominant ISR contributions to the second jet render 
the FSR contributions less clear, 
we can see that the large $\mu$ dependences observed in the separate processes 
are compensated to yield a stable distribution in the summed result, 
implying good matching in the FSR.

Despite an overall compensation,
a small $\mu$ dependence is left in the summed result 
in Fig.~\ref{fig:matching2}, near the lower limit of $\Delta R$.
This is due to the multiple radiation effects in PS.
Our matching method takes care about the leading contribution of PS only.
The method is tailored so as to achieve matching between this leading 
contribution and the divergent terms in radiative processes.
The matching may be deteriorated by the existence of additional jets 
in PS in the $Q^{2}$ range where $\mu^{2}$ is moved.
On the other hand, 
only one photon is assumed in the present study because the contribution from 
multiple photon radiation is expected to be negligible.
This must be the reason why we observe almost perfect compensation 
in Fig.~\ref{fig:matching1}.

\begin{figure}[tp]
\begin{center}
\includegraphics[scale=0.6]{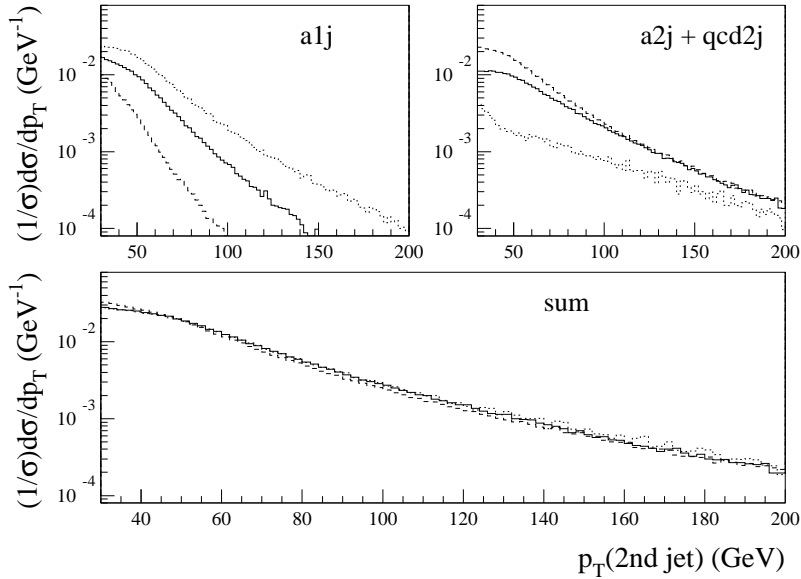}
\caption{\label{fig:matching3}
$p_{T}$ distribution of the second jet.
The distributions for $a1j$ and $a2j + qcd2j$ are shown separately 
in the upper panels, and the sum is shown in the lower panel.
The results for $\mu = \mu_{0}/2$, $\mu_{0}$, and $2\mu_{0}$ are 
represented by dashed, solid, and dotted histograms, respectively.
}
\end{center}
\end{figure}

The matching in the initial-state QCD radiation can be tested by 
investigating the second-jet $p_{T}$ distribution.
The $a1j$ and $a2j + qcd2j$ results are shown separately in the upper panels 
in Fig.~\ref{fig:matching3}, and the sum is shown in the lower panel.
We again observe a strong compensation for the $\mu$ dependence in this result, 
implying good matching in the initial-state QCD radiation.
However, some $\mu$ dependence remains in the summed result, 
although this may be difficult to see in Fig.~\ref{fig:matching3} 
where the cross sections are displayed on a logarithmic scale.
The reason for the remaining dependence must be identical to that discussed 
for the results shown in Fig.~\ref{fig:matching2}.

In conclusion, 
we have observed a strong compensation between separate processes for 
the energy-scale dependence of the observable quantity distributions, 
which are sensitive to the matching in the QED final-state and 
QCD final- and initial-state radiations.
These observations imply good matching between photon and jet radiation 
in PS and those in MEs of radiative processes.
We can obtain stable distributions by adding the results of all processes.
The remaining energy-scale dependence can be attributed to 
a multiple-radiation effect in PS.

\section{Comparison with measurements at LHC}
\label{sec:comparison}

The performance of the developed event generator is examined in this section 
by comparing the simulation results with actual measurement data obtained 
at the LHC.
Because a large amount of measurement data is currently available, 
we compare our results with selected measurements below.

\subsection{ATLAS measurement (1)}
\label{sec:atlas1}

\begin{figure}[tp]
\begin{center}
\includegraphics[scale=0.6]{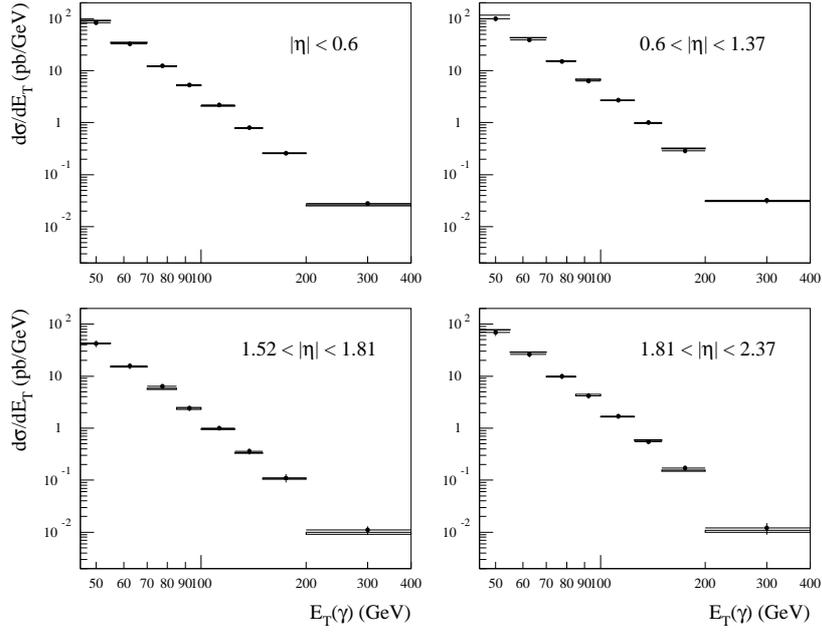}
\caption{\label{fig:atlas1a}
Comparison with the ATLAS measurement data reported in~\cite{Aad:2011tw}.
The simulation results are represented by boxes.
The vertical sizes of the boxes indicate the statistical error 
of the simulation.
The measurement results are shown with plots having error bars,  
which indicate the total error quoted in Tables A.1 - A.4 
in the ATLAS paper, 
where the luminosity uncertainty of 3.4\% is not included.
}
\end{center}
\end{figure}

\begin{figure}[tp]
\begin{center}
\includegraphics[scale=0.6]{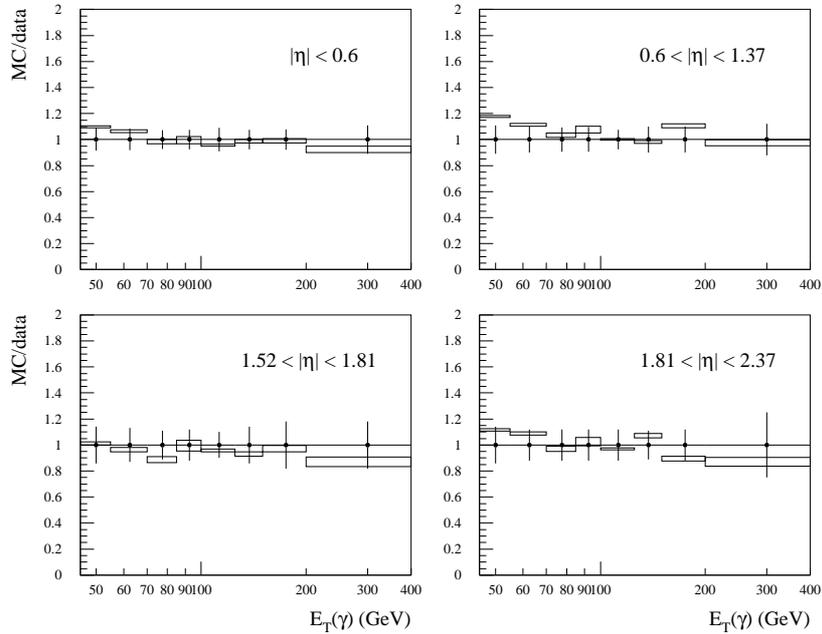}
\caption{\label{fig:atlas1b}
Ratio of the simulation results to the ATLAS measurement data 
reported in~\cite{Aad:2011tw}.
The notation is identical to that in Fig.~\ref{fig:atlas1a}.
}
\end{center}
\end{figure}

We neglect to consider the first CMS~\cite{Khachatryan:2010fm} and 
ATLAS~\cite{Aad:2010sp} measurements because they are based on 
limited data obtained at an early stage of these experiments.
In this subsection, we compare our simulation results with the second 
ATLAS measurement~\cite{Aad:2011tw} based on 35-pb$^{-1}$ data 
obtained at a CM energy of 7 TeV and collected in 2010.
They measured the photon $E_{T}$ ($= p_{T}$) spectra in the range of 
45 $< E_{T} < $ 400 GeV in four pseudorapidity ($\eta$) regions: 
$|\eta| < 0.6$, $0.6 < |\eta| < 1.37$, $1.52 < |\eta| < 1.8$, 
and $1.8 < |\eta| < 2.37$.
They applied the cone isolation condition in Eq.~(\ref{eq:iso}) 
with $E_{T}^{\rm iso}$ = 3 GeV at the detector level, and estimated 
that this condition is equivalent to $E_{T}^{\rm iso}$ = 4 GeV 
with $\Delta R = 0.4$ at the particle level, 
in which all stable particles except for neutrinos and muons are included.

Following the measurement conditions, 
we simulated the events in 7-TeV proton-proton collisions, 
with the conditions $E_{T} > $ 45 GeV, 
$|\eta| < 2.37$, and $E_{T}^{\rm cone} < $ 4 GeV, 
and binned them to derive the differential cross section. 
A modified leading-order PDF, MRST2007LO*~\cite{Sherstnev:2007nd}, 
was used for the hard-interaction generation because 
this PDF is widely used in experiments for MC simulations.
The generation conditions and preselection conditions for 
the events to be fed to PYTHIA were sufficiently relaxed 
in order not to bias the simulation results.
Effectively, no constraint was applied to associated jets.

The simulation results are compared with the measurement data 
in Fig.~\ref{fig:atlas1a}, 
and the ratio between them is presented in Fig.~\ref{fig:atlas1b}.
Because the measurement error is dominated by systematic errors, 
the simulation was carried out separately for low-$E_{T}$ ($< 100$ GeV) and
high-$E_{T}$ ($> 100$ GeV) regions in order to save the computational cost.
Although there is no guarantee of the overall normalization 
for simulations based on tree-level MEs, 
the simulation results are in good agreement with the measurement data within 
the quoted measurement errors, without any adjustment of the normalization.

Despite an overall agreement, 
we can observe a slight enhancement of the simulation in low-$E_{T}$ regions.
The energy scale $\mu$ is another free parameter in the simulation 
that we can optimize.
However, a change of $\mu$ does not result in significant alteration 
of the $E_{T}$ dependence, although it alters the overall yield.
Even if we change the $\mu$ value from $\mu_{0}/2$ to $2\mu_{0}$, 
as in the study of the matching in Sec.~\ref{sec:matching}, 
the variation in the normalized cross section, $(1/\sigma)d\sigma/dE_{T}$, 
is confined within $\pm 10\%$ from the $\mu = \mu_{0}$ result.
The use of a leading-order PDF, CTEQ6L1~\cite{Pumplin:2002vw}, 
reduces the overall yield by approximately 10\%, 
but it does not significantly alter the $E_{T}$ dependence.
This must be simply due to the lower gluon density in this PDF.

\subsection{CMS measurement (1)}
\label{sec:cms1}

\begin{figure}[tp]
\begin{center}
\includegraphics[scale=0.6]{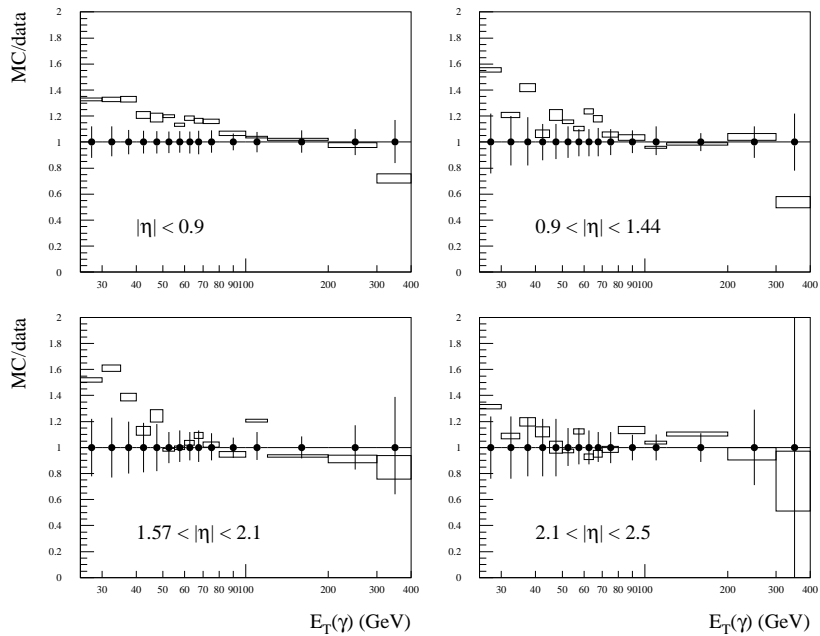}
\caption{\label{fig:cms1}
Comparison with the CMS measurement data for inclusive direct-photon 
cross section reported in~\cite{Chatrchyan:2011ue}.
The ratio of the simulation to the measurement is illustrated using boxes.
The vertical size of each box represents the statistical error 
of the simulation.
The error bars indicate the quadratic sum of the statistical and systematic 
errors of the measurement presented in Table 6 of the CMS paper, 
where the luminosity uncertainty of 4.0\% is not included.
}
\end{center}
\end{figure}

The inclusive direct-photon cross section was also measured by 
CMS~\cite{Chatrchyan:2011ue} based on 36-pb$^{-1}$ data collected in 2010.
They measured the differential cross section, $d^{2}\sigma/dE_{T}/d\eta$, 
in the $E_{T}$ range of 25 $< E_{T} <$ 400 GeV 
in four pseudorapidity ranges within $|\eta| <$ 2.5.
Although an isolation condition is required in the measurement, 
the signal definition is not explicitly described.
As they apply the condition, $E_{T}^{\rm cone} <$ 5 GeV with $\Delta R = 0.4$, 
to the simulated events in their discussions, 
we assume this condition to be the signal definition in the present study.

We carried out a simulation by appropriately setting the $E_{T}$ threshold 
and the $\eta$ range, 
and applied the above isolation condition to the simulated events.
The other simulation conditions are identical to those 
in Sec.~\ref{sec:atlas1}.
The $E_{T}^{\rm cone}$ was evaluated using all stable particles except 
for neutrinos and muons, although this detail is not given 
in the CMS paper.
Again, in order to save the computational cost, 
we performed the simulation in three separate $E_{T}$ ranges: 
$25 < E_{T} < 50$ GeV, $50 < E_{T} < 100$ GeV, and $100 < E_{T} < 400$ GeV.

We present the ratio between the simulation and measurement 
in Fig.~\ref{fig:cms1}.
Although the simulation is in good agreement with the measurements 
at high $E_{T}$ ($\gtrsim 100$ GeV), 
a significant excess of the simulation is observed in lower-$E_{T}$ regions.
This tendency is consistent with the observation in Sec.~\ref{sec:atlas1}.
CMS compared their results with an NLO prediction 
by JETPHOX~\cite{Catani:2002ny} 
and observed a similar excess of the prediction at low $E_{T}$.
However, the excess that they observed is smaller than the present one.
As we will show later, 
the 2-jet contributions ($a2j + qcd2j$) increase as $E_{T}$ decreases.
Hence, the contribution of further higher-order processes may also be large 
at low $E_{T}$.
JETPHOX partly involves 3-jet contributions as an NLO correction to 
the fragmentation processes.
The observed excess of the simulation may be related to the contribution 
from missing higher-order processes.

\subsection{ATLAS measurement (2)}
\label{sec:atlas2}

\begin{figure}[tp]
\begin{center}
\includegraphics[scale=1.0]{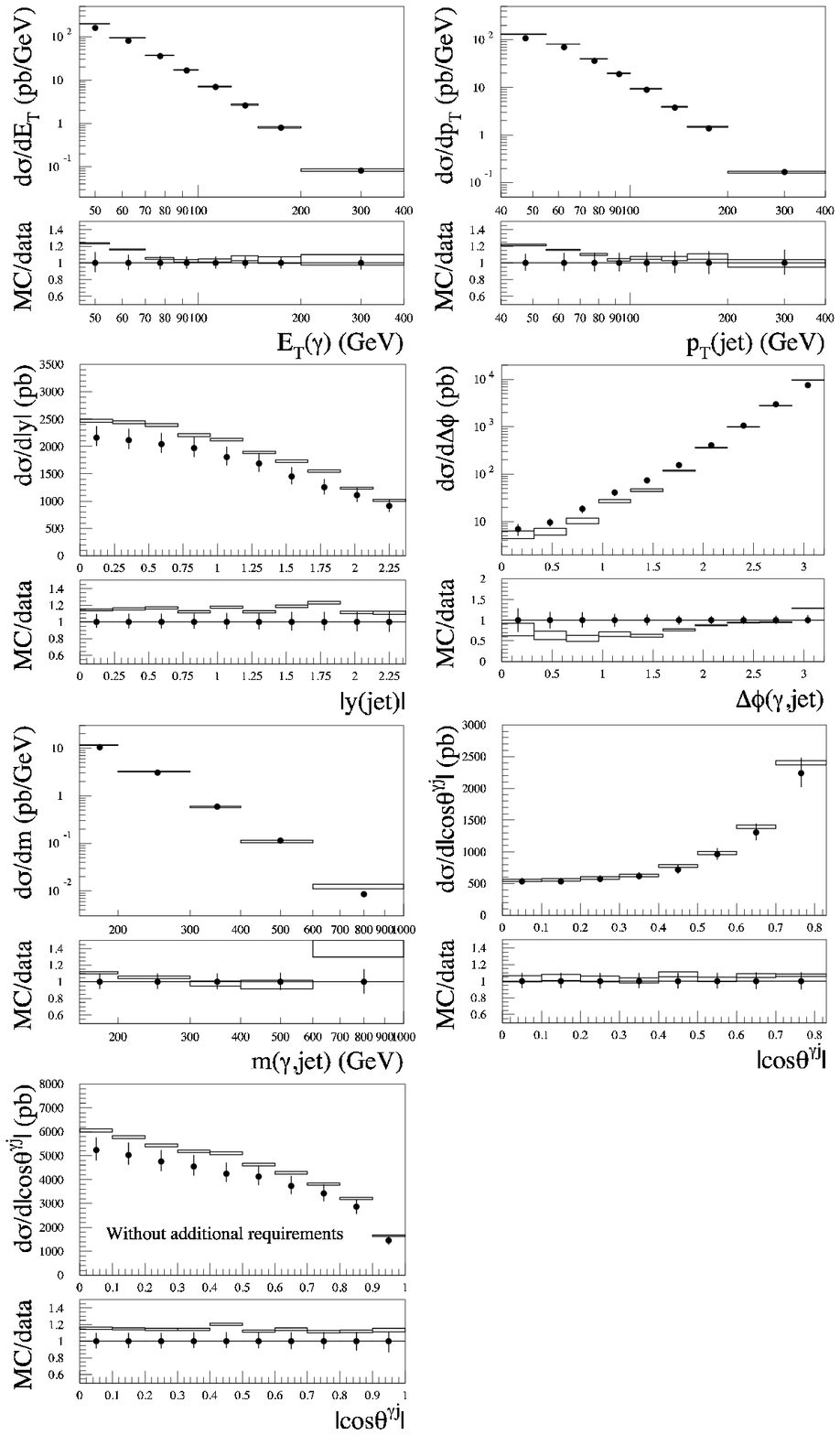}
\caption{\label{fig:atlas2}
Comparison with the ATLAS measurements for $\gamma$ + jet events 
reported in~\cite{Aad:2013gaa}.
The simulation results are represented by boxes and 
the measurement results are plotted with error bars.
The luminosity uncertainty of 3.4\% is included in the error bars.
}
\end{center}
\end{figure}

ATLAS measured the properties of photons, jets, and their correlations 
in $\gamma$ + jet events~\cite{Aad:2013gaa}.
The measurement was based on the same data as the measurement discussed 
in Sec.~\ref{sec:atlas1}, with the same photon selection condition.
In order to study the jet properties, 
they reconstructed hadron jets using the anti-$k_{T}$ clustering 
with $R$ = 0.6, and applied the conditions,
\begin{equation}\label{eq:jet-atlas2}
  p_{T}({\rm jet}) > 40 {\rm ~GeV}/c ,  ~~|\eta({\rm jet})| < 2.37 ,  
  ~~\Delta R(\gamma, {\rm jet}) > 1.0 ,
\end{equation}
to the reconstructed jets.
The notation for these parameters is identical to that 
in Sec.~\ref{sec:matching}.

In order to simulate the measurement results, 
we applied the same jet clustering to the simulation events used 
in Sec.~\ref{sec:atlas1} employing FastJet 3.0.3, 
and also applied the kinematical condition in Eq.~(\ref{eq:jet-atlas2}) 
to the reconstructed jets.
All stable particles in the pseudorapidity range of $|\eta| < 5.0$, 
including neutrinos, were used for the clustering.
The observation of at least one such jet was required, 
and the leading (highest-$p_{T}$) jet was taken as the jet to be studied.
The simulation results are compared with the measurements 
in Fig.~\ref{fig:atlas2}.
The presented results correspond to those in Figs.~9 - 14 and 17, 
and Tables 1 - 7 in the ATLAS paper~\cite{Aad:2013gaa}.
The plotted error bars represent the quadratic sums of the statistical and 
systematic errors listed in the tables.

The agreement between the simulation and measurement is marginal.
The majority of the discrepancies that we observe in Fig.~\ref{fig:atlas2} 
are caused by the small excess of the simulation results 
at low $E_{T}(\gamma)$ and low $p_{T}({\rm jet})$.
We observe nearly complete agreement for all results, 
excluding the $\Delta\phi(\gamma,{\rm jet})$ distribution, 
if we assume an additional 20\% inefficiency around $p_{T}$(jet) = 50 GeV/$c$ 
for the jet detection, 
or 10\% photon inefficiency plus 10\% jet inefficiency for $E_{T}(\gamma)$ 
and $p_{T}({\rm jet})$ around 50 GeV.

The discrepancy in the $\Delta\phi(\gamma,{\rm jet})$ distribution, {\it i.e.}, 
the distribution of the azimuthal angle separation between the photon and jet, 
cannot be significantly reduced by assuming additional inefficiencies.
As described in Sec.~\ref{sec:strategy}, we deactivated the underlying-event 
simulation in PYTHIA. 
Angular distributions are sometimes very sensitive to such soft interactions.
However, we observed no significant improvement when the underlying-event 
simulation was activated.
Rather, this simply reduced the overall yield because of 
an additional contribution to $E_{T}^{\rm cone}$ for the photon isolation.

A similar discrepancy in the $\Delta\phi(\gamma,{\rm jet})$ distribution 
was observed by ATLAS when their measurement result was compared 
with predictions from JETPHOX and HERWIG, 
while the SHERPA and PYTHIA predictions exhibited better agreement 
(see Fig.~12 in~\cite{Aad:2013gaa}).
SHERPA includes multi-jet ME contributions, 
and PYTHIA simulates multiple radiation effects in the collinear 
approximation by increasing the PS energy scale.
Hence, the discrepancy in $\Delta\phi(\gamma,{\rm jet})$ observed 
in the present study must be due to the lack of multi-jet 
(three or more jet) contributions in our simulation.

\subsection{CMS measurement (2)}
\label{sec:cms2}

\begin{figure}[tp]
\begin{center}
\includegraphics[scale=0.7]{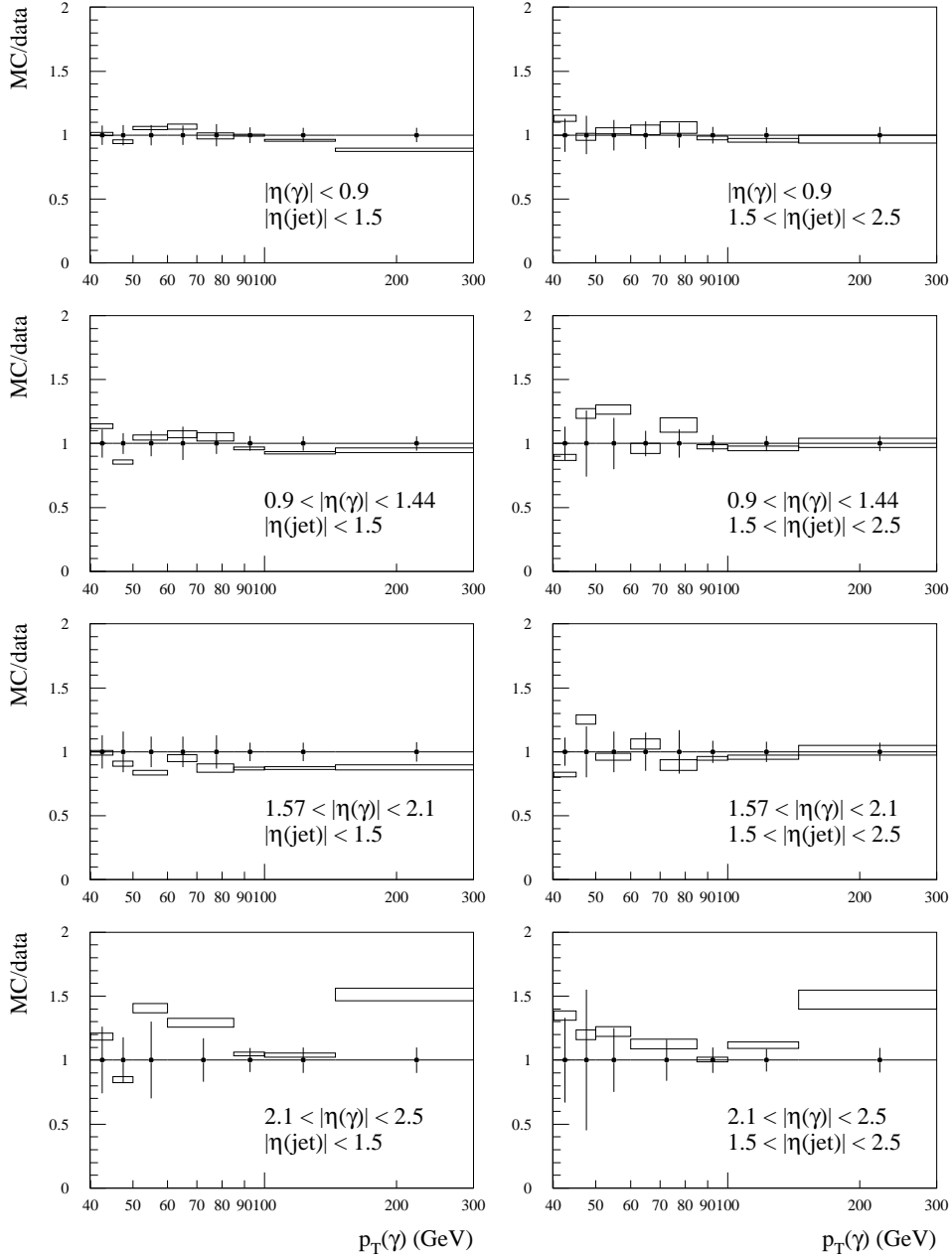}
\caption{\label{fig:cms2}
Comparison with the CMS measurement data for $\gamma$ + jet events 
reported in~\cite{Chatrchyan:2013mwa}.
The ratio between the simulation and measurement concerning the 
triple-differential cross section is presented as a function of 
the photon $p_{T}$.
The results are presented in the eight angular regions, 
four $|\eta(\gamma)|$ $\times$ two $|\eta({\rm jet})|$, 
shown in the figure.
The error bars include the luminosity uncertainty of 2.2\%.
}
\end{center}
\end{figure}

CMS also measured the properties of $\gamma$ + jet events in terms of the 
triple-differential cross section, 
$d^{3}\sigma/(dp_{T}^{\gamma}d\eta_{\gamma}d\eta_{\rm jet})$, 
using 2.14-fb$^{-1}$ data for 7-TeV collisions~\cite{Chatrchyan:2013mwa}, 
in the pseudorapidity range of $|\eta| < 2.5$ for both photons and jets, 
and in the $p_{T}$ ranges of $40 < p_{T} < 300$ GeV/$c$ for photons and 
$p_{T} > 30$ GeV/$c$ for jets.
The hadron jets were reconstructed using the anti-$k_{T}$ clustering 
with $R$ = 0.5, 
and the separation between the photon and jet was required to be 
$\Delta R(\gamma,{\rm jet}) > 0.5$.

The simulation was carried out following the above conditions 
by using the MRST2007LO* PDF.
Although the isolation condition for the photon signal is again not 
explicitly defined, we assumed the condition 
$E_{T}^{\rm cone}(\Delta R = 0.4) < $ 5 GeV in our simulation 
because CMS required this condition in their simulation.
In order to save the computational cost, 
the simulation was carried out separately for the $p_{T}(\gamma)$ ranges 
below and above 85 GeV/$c$.

The ratio of the simulation to the measurement data is presented 
in Fig.~\ref{fig:cms2}.
We observe reasonable agreement between them; 
at least, no remarkable systematic discrepancy is observed.
A marginal observed discrepancy is a systematic deficit of the simulation 
in the angular region of 1.57 $< |\eta(\gamma)| <$ 2.1 
and $|\eta({\rm jet})| <$ 1.5. 
However, a deficit is observed in the normalization only and 
the discrepancy is marginal compared to the measurement errors.
We also observe a marked discrepancy in the highest-$p_{T}(\gamma)$ bins 
in the largest-$|\eta(\gamma)|$ regions. 
This discrepancy was also observed by CMS in comparison with 
the JETPHOX prediction in their report.
We have no comment on this observation.
We must note that, in this result, we do not observe the excess 
in the simulation result at low $p_{T}$ that we observed 
in Sec.~\ref{sec:atlas2}.

\subsection{ATLAS measurement (3)}
\label{sec:atlas3}

\begin{figure}[tp]
\begin{center}
\includegraphics[scale=0.7]{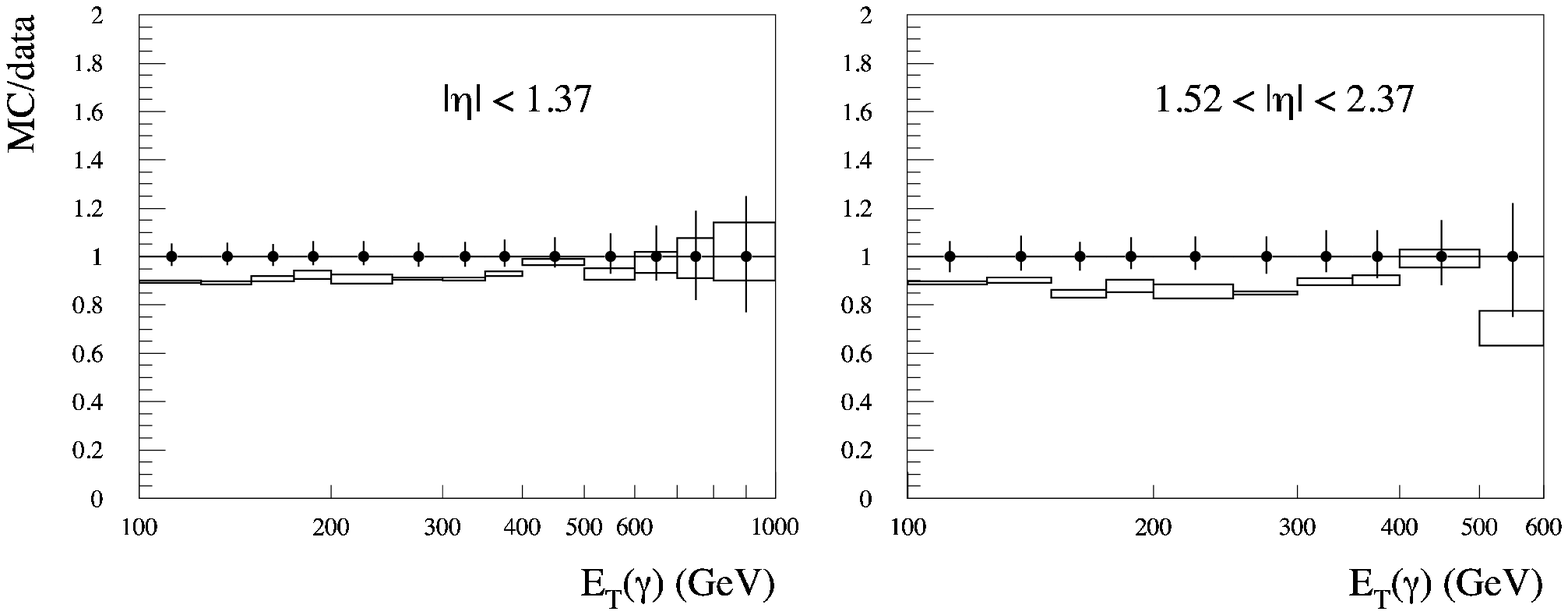}
\caption{\label{fig:atlas3}
Comparison with the ATLAS measurement data for high-$E_{T}(\gamma)$ 
regions published in~\cite{Aad:2013zba}.
The ratio of the simulation to the measurement concerning 
the inclusive cross section is represented by boxes 
as a function of $E_{T}(\gamma)$ in two angular regions.
The error bars indicate the quadratic sum of the statistical and systematic 
errors of the measurement, 
including the luminosity uncertainty of 1.8\%.
}
\end{center}
\end{figure}

ATLAS published inclusive $E_{T}(\gamma)$ distribution results 
in high-$E_{T}$ regions 
using 4.6-fb$^{-1}$ data collected in 2011~\cite{Aad:2013zba}.
They presented the differential cross section, $d\sigma/dE_{T}$, 
in the $E_{T}$ range of 100 GeV $< E_{T} <$ 1 TeV
in two pseudorapidity ranges, $|\eta| < 1.37$ and $1.52 < |\eta| < 2.37$.
The definition of the isolation condition for the signal photons was 
explicitly given as $E_{T}^{\rm cone}(\Delta R = 0.4) <$ 7 GeV, 
where $E_{T}^{\rm cone}$ was evaluated by using all stable particles 
included in the cone, excluding neutrinos and muons.
The simulation was carried out following these measurement conditions 
using the MRST2007LO* PDF.
The events were simulated separately for $E_{T}$ ranges below and 
above 250 GeV.

The ratio of the simulation results to the measurement data is presented 
in Fig.~\ref{fig:atlas3}.
Contrary to the previous comparison results, 
the simulation exhibits a systematic deficit of approximately 10\% 
through the measurement range.
We observe very good agreement with the measurement if we increase 
the overall normalization of the simulation by 10\%.
As nothing is essentially changed in the simulation, 
this discrepancy indicates an inconsistency between the measurements.
However, since the error of the measurements discussed previously was 
at the level of 10\% or larger, 
this inconsistency could be considered as a tolerance within the 
measurement errors.

\subsection{Summary of the comparison}
\label{sec:comp-discuss}

Overall, the simulation based on the developed MC event generator 
reproduces direct-photon measurement results at the LHC reasonably well, 
not only for inclusive photon spectra but also for the properties 
of associated jet production. 
Here, we emphasize that this {\it jet} is not an energetic parton, 
but is a realistic hadron jet reconstructed from simulated stable particles.

Despite an overall consistency, some discrepancies seem to exist 
in low-$p_{T}$ regions ($\lesssim 50$ GeV/$c$). 
This observation may indicate the necessity for further higher-order 
processes including three or more jets in the final state.
However, this observation is not conclusive as the measurements are not 
fully consistent with each other.
New measurements or reanalyses are necessary for confirmation.

As we have mentioned in the introduction, 
the signal events for which the measurement results are presented 
must be unambiguously defined 
to permit rigorous comparisons with theoretical predictions.
However, the definition of the photon isolation condition is not explicitly 
presented in some measurement reports.
We would like to find clear descriptions in future reports.
In addition, we are concerned about the estimation of 
the photon-isolation efficiency.
The efficiency is evaluated in the majority of the measurements 
using the PYTHIA simulation.
As we have noted in a previous study~\cite{Odaka:2012ry}, 
the photon-radiation probability is too small in the PYTHIA 6.4 
PS simulation. 
We are concerned that this problem may have somewhat affected 
the measurement results published to date.

\section{Characteristic properties}
\label{sec:discussion}

\begin{figure}[tp]
\begin{center}
\includegraphics[scale=0.6]{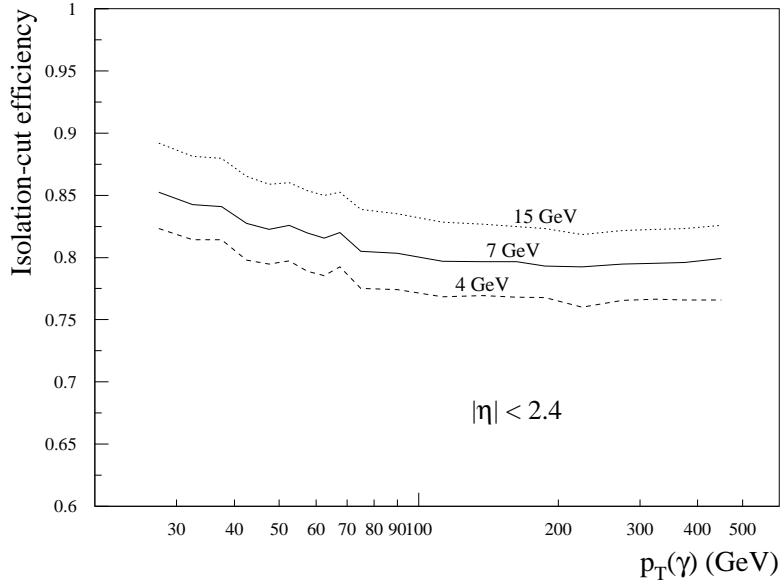}
\caption{\label{fig:isoeff}
Efficiency of the photon-isolation cut.
The ratio between the cross sections with and without the cut is displayed 
as a function of the photon $p_{T}$.
The simulation results under the 7-TeV LHC condition are presented for 
three cut values of $E_{T}^{{\rm iso}}$ = 4, 7, and 15 GeV.
}
\end{center}
\end{figure}

In this section, we discuss the characteristic properties of the direct-photon 
production process in hadron collisions using our simulation results.
We first examine the impact of the photon-isolation requirement.
We can evaluate the cross section without applying the isolation cut, 
as the cross section is made finite by summing all-order contributions 
employing PS and FF in collinear regions.
Figure~\ref{fig:isoeff} shows the efficiency of the isolation cut, {\it i.e.}, 
the ratio between the cross sections with and without the cut. 
This simulation was carried out under the 7-TeV LHC condition 
using the MRST2007LO* PDF with our standard energy-scale setting.
The direct photon was required to be produced in the angular region of 
$|\eta| < 2.4$.

The efficiency was evaluated for three values of the simple cone-$E_{T}$ cut 
defined in Eq.~(\ref{eq:iso}) with $\Delta R = 0.4$ 
as a function of the photon $p_{T}$.
It can be seen that the efficiency is approximately 80\% and 
decreases slightly as $p_{T}(\gamma)$ increases.
It almost saturates at high $p_{T}(\gamma)$ ($\gtrsim 200$ GeV).
The dependence on the $E_{T}$-cut value is moderate in the examined range.
The simulation shows that approximately one half of the $qcd2j$ events are 
rejected by the cut.
On the other hand, 
only a few percent of the $a1j$ and $a2j$ events are rejected, 
because these processes do not have any enhancement 
in photon-quark collinear regions.
Although the rejection by the isolation cut is not very large, 
it is inadvisable to correct the measured cross section 
for this inefficiency, 
because substantial uncertainties are suspected in the efficiency evaluation, 
such as uncertainties in non-perturbative effects and PS-model dependencies.

\begin{figure}[tp]
\begin{center}
\includegraphics[scale=0.6]{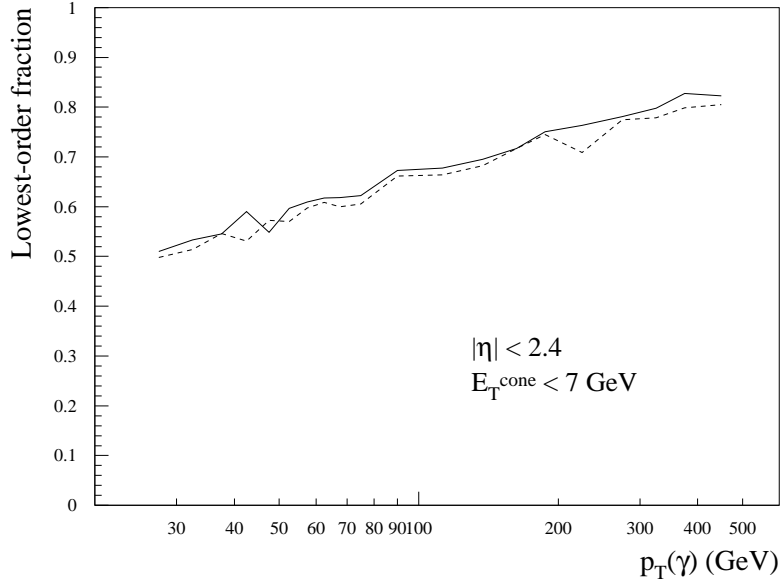}
\caption{\label{fig:lofrac}
Fraction of the lowest-order $\gamma$ + 1-jet contribution 
after the isolation cut of $E_{T}^{{\rm cone}} <$ 7 GeV 
with our standard energy-scale setting.
The ratio of the $\gamma$ + 1-jet cross section without the soft-gluon 
correction to the summed cross section, $a1j + a2j + qcd2j$, is displayed 
as a function of the photon $p_{T}$.
The simulation results obtained by using two different PDFs, 
MRST2007LO* (solid) and CTEQ6L1 (dashed), are presented.
}
\end{center}
\end{figure}

Figure~\ref{fig:lofrac} shows the fractional contribution of the lowest-order 
$\gamma$ + 1-jet production processes after the isolation cut 
of $E_{T}^{\rm cone} <$ 7 GeV.
Here, we evaluated the ratio of the $\gamma$ + 1-jet cross section 
without applying the soft-gluon correction 
with respect to the summed cross section.
The result is presented as a function of $p_{T}(\gamma)$.
The soft-gluon correction reduces the $\gamma$ + 1-jet cross section, 
but the reduction is not larger than 10\% through the examined 
$p_{T}(\gamma)$ range.
We can see that the fraction is only approximately 50\% 
at $p_{T}(\gamma)$ = 30 GeV, 
despite the fact that a large portion of $qcd2j$ events have been rejected 
by the isolation cut.
In other words, as we have emphasized, the correction by the $\gamma$ + 2-jet 
contributions is quite large, being nearly 100\%.
Hence, we suspect that further higher-order contributions may also be large 
at low $p_{T}(\gamma)$.
In Fig.~\ref{fig:lofrac}, the results of the simulations conducted 
using two PDFs, MRST2007LO* and CTEQ6L1, are presented.
The difference between the two results is only a few percent.

Note that the result in Fig.~\ref{fig:lofrac} is not physically meaningful.
The sharing between the processes depends on the choice of energy scales.
However, as we adopted a natural choice of energy scales,
the above findings must be helpful for furthering understanding 
of direct-photon production.
In theoretical calculations, the choice of $\mu = p_{T}(\gamma)/2$ is also 
frequently adopted.
The lowest-order fraction is further reduced for this choice.

\begin{figure}[tp]
\begin{center}
\includegraphics[scale=0.6]{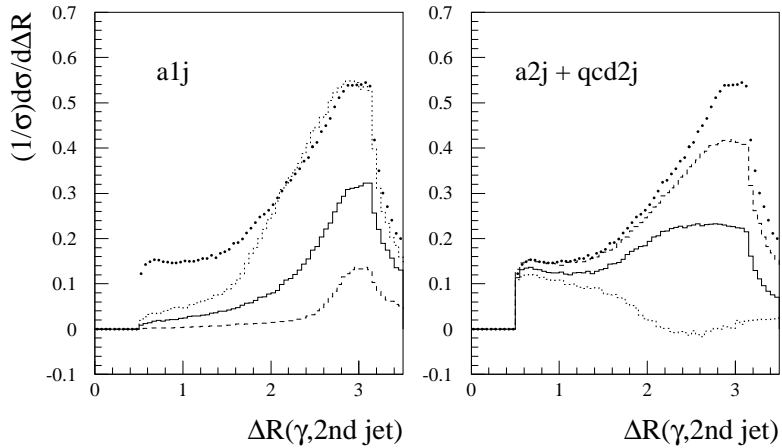}
\caption{\label{fig:drgamj2}
Distribution of the $\Delta R$ separation between the photon and second jet.
The results for $a1j$ and $a2j + qcd2j$ are shown separately.
Three results for energy scales of $\mu = \mu_{0}/2$ (dashed), 
$\mu = \mu_{0}$ (solid), and $\mu = 2\mu_{0}$ (dotted) are presented.
The summed distribution with $\mu = \mu_{0}$ is illustrated using thick dots 
for comparison.
}
\end{center}
\end{figure}

As we have shown in Sec.~\ref{sec:matching},
the final-state QED divergence produces a flat distribution of events 
in the distribution of the $\Delta R$ separation between the photon and 
second jet, $\Delta R(\gamma,{\rm 2nd~jet}$), 
at small $\Delta R$ ($\lesssim 1.5$).
The existence of final-state divergences is characteristic in 
higher-order processes involving two or more jets.
The contributions of the lowest-order 1-jet ($a1j$) and higher-order 2-jet  
($a2j + qcd2j$) processes to this $\Delta R$ distribution are shown separately 
in Fig.~\ref{fig:drgamj2}.
We used the simulation sample for the matching test in Sec.~\ref{sec:matching} 
and applied the same analyses to the events.
The only difference is in the separation of the processes.

As in Sec.~\ref{sec:matching},
the results for the energy-scale choices of $\mu = \mu_{0}/2$ and 
$\mu = 2\mu_{0}$ are shown together with those for the standard 
choice of $\mu = \mu_{0}$.
In this result, we observe a very large energy-scale dependence 
in large-$\Delta R$ regions.
However, the dependence reduces as $\Delta R$ decreases, 
and the flat distribution at $\Delta R \lesssim 1.5$ cannot be reproduced 
by the lowest-order $a1j$ contribution.
Thus, if we can measure this distribution at small $\Delta R$, 
this will be a direct quantitative measurement of higher-order contributions 
that necessarily have QED divergences in the final state.

\begin{figure}[tp]
\begin{center}
\includegraphics[scale=0.6]{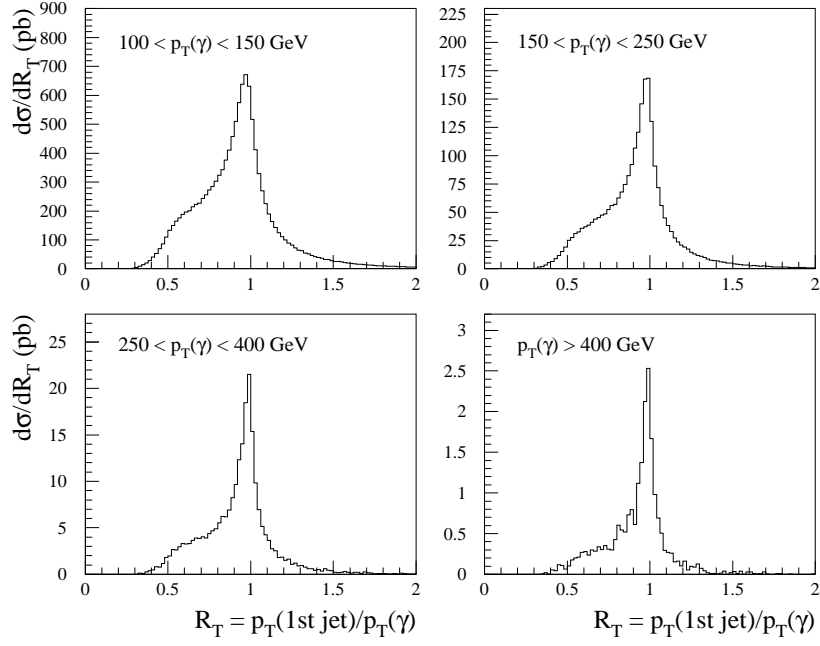}
\caption{\label{fig:ptcor1}
Distribution of the $p_{T}$ ratio between the photon and leading jet, 
$R_{T} = p_{T}({\rm 1st~jet})/p_{T}(\gamma)$.
The simulation sample used in the matching test in Sec.~\ref{sec:matching} 
was used here.
The photon and jets were selected according to the conditions 
in Eqs.~(\ref{eq:photon1}), (\ref{eq:iso1}), and (\ref{eq:jet1}), 
requiring the detection of at least one jet.
The results for four $p_{T}(\gamma)$ ranges are presented.
}
\end{center}
\end{figure}

\begin{figure}[tp]
\begin{center}
\includegraphics[scale=0.6]{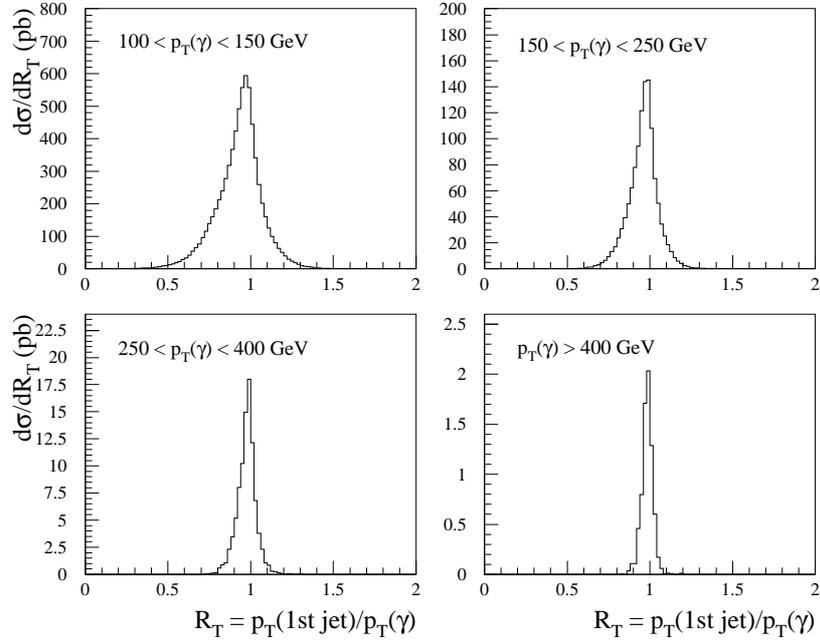}
\caption{\label{fig:ptcor2}
Identical to Fig.~\ref{fig:ptcor1}, but the number of detected jets is 
required to be exactly one.
}
\end{center}
\end{figure}

Finally, we examine the correlation between the photon and the leading jet.
As the direct-photon production is a two-body $\gamma$ + jet production 
at the lowest order, this process is considered to constitute 
a good sample for jet-energy calibration; 
the $p_{T}$ of hadron jets can be calibrated by referring to the photon 
$p_{T}$ which can be measured precisely.
However, higher-order processes may deteriorate this capability.
The distribution of the ratio between the photon $p_{T}$ and the leading-jet 
$p_{T}$, $R_{T} = p_{T}({\rm 1st~jet})/p_{T}(\gamma)$, 
in the events used for the matching test discussed in Sec.~\ref{sec:matching} 
are shown in Fig.~\ref{fig:ptcor1}.
The photon and jet were selected according to the conditions 
in Eqs.~(\ref{eq:photon1}), (\ref{eq:iso1}), and (\ref{eq:jet1}), 
and at least one jet was required to be detected.
As a result of the very loose condition for jet detection, 
at least one jet is almost always detected in photon-detected events;
the jet-detection efficiency is more than 99\%.

The $R_{T}$ distribution is shown separately for four $p_{T}(\gamma)$ ranges 
in the figure.
Although a concentration of events is observed around $R_{T}$ = 1, 
we observe a shoulder-like structure at small $R_{T}$.
This structure is created by the contribution of second jets 
produced in association with the leading jet.
Thus, it is expected that this structure can be eliminated by suppressing 
the second-jet contribution.
As an example, we required that the number of detected jets 
was exactly one.
Approximately one half of the events were rejected by this requirement.
The loose jet-detection condition became a tight veto condition 
in this part of the study.
Figure~\ref{fig:ptcor2} shows the results obtained under this requirement, 
and it is apparent that the structure at small $R_{T}$ has vanished.
In addition, the long tail to the $R_{T} > 1$ side has been eliminated.
This is because this tail was predominantly produced by $qcd2j$ contributions, 
in which a detectable jet was produced in association with the photon.

The $R_{T}$ distributions in Fig.~\ref{fig:ptcor2} still have finite widths 
because of the small-angle undetectable jets and soft QCD interactions.
In addition, especially in low $p_{T}(\gamma)$ ranges, the distribution has 
a longer tail to the $R_{T} < 1$ side.
This tail is due to the particles escaping from the jet reconstruction.
The peak position is somewhat shifted from unity, 
possibly as a result of this effect.

These observations imply that direct-photon production provides a good data 
sample for the calibration of hadron-jet momenta 
if we can efficiently eliminate the contributions from additional jets.
In practical applications, we must account for the shift in the peak 
position caused by the unavoidable escape of particles from jets.
The shift would be larger than that in Fig.~\ref{fig:ptcor2} 
in real-world cases, because of the additional detection resolution.

\section{Conclusion}
\label{sec:conclusion}

We have developed an event generator for direct-photon production in hadron 
collisions, consistently combining $\gamma$ + 2-jet production processes 
with the lowest-order $\gamma$ + 1-jet production using a subtraction method.
The final-state QED collinear divergences are also subtracted from 
$\gamma$ + 2-jet MEs.
The subtracted QED components are restored by combining QCD 2-jet production 
events in which an energetic photon is produced by a QCD/QED-mixed PS and 
a simulation employing an FF.
The photon radiation is enforced in this PS/FF simulation to facilitate 
efficient event generation.
The generated events can be passed to PYTHIA in order to add hadronization 
and decay simulations.
We can conduct realistic simulations of photon isolation and hadron-jet 
clustering using the obtained particle-level event information.

The self-consistency of the developed event generator has been examined 
by investigating observable quantities in the simulated events.
The obtained distributions, which are sensitive to the matching between 
$\gamma$ + 2-jet production and other contributions, 
exhibit good stability against variation of energy scales, 
implying good matching between the included PS and ME-based simulations 
for photon and jet radiation.

The performance of the simulation has been tested by comparing the simulation 
results with measurement data at the LHC.
The comparison was carried out concerning not only the inclusive 
properties of the photon, but also the properties of the associated hadron jet.
The simulation results show reasonable agreement with the measurement data, 
especially in high photon-$p_{T}$ regions, 
where $p_{T}(\gamma) \gtrsim 100$ GeV.
Although substantial discrepancies have been found in low-$p_{T}$ regions of 
the photon and jet, {\it i.e.}, $p_{T} \lesssim 50$ GeV, 
the observations are not conclusive 
because the measurement results are not fully consistent with each other.

We have demonstrated some characteristic properties of direct-photon production 
using the simulated events.
The efficiency of the photon-isolation cut is approximately 80\% 
if a simple cone-$E_{T}$ cut is applied with cut values 
of approximately 10 GeV.
Although the cut predominantly rejects final-state QED radiation events, 
the contribution of the 2-jet processes is very large 
even after the isolation cut.
The lowest-order $\gamma$ + 1-jet contribution is as small as 50\% 
around $p_{T}(\gamma)$ = 50 GeV with our standard energy-scale setting. 
This fraction gradually increases as $p_{T}(\gamma)$ increases.
Hence, discrepancies between the simulation and measurement observed 
at low $p_{T}$ may indicate the necessity for the incorporation of 
further higher-order processes, 
including the production of three or more jets.
The higher-order contributions can be quantitatively 
measured by investigating the angular separation between the photon and 
the second jet.
We have also shown that the direct-photon process provides a good sample 
for the jet-momentum calibration, if the second-jet contribution can be 
effectively eliminated.

As we have noted in the introduction, 
it is important to give an unambiguous definition of the photon isolation 
for the signal events in measurements, 
in order to facilitate rigorous comparison with theoretical predictions.
The isolation must be defined using the event information 
at the particle (hadron) level.
We could not find such a definition for some of the measurements 
that we used for the comparison conducted in the present study.
Future measurements are desired to present the definition explicitly.

\section*{Acknowledgments}

This work has been carried out as an activity of the NLO Working Group, 
a collaboration between the Japanese ATLAS group and the numerical analysis 
group (Minami-Tateya group) at KEK.
The authors wish to acknowledge useful discussions with the group members.

% References
%\bibliography{direct.bib}
\include{direct_bbl}

\end{document}

%% file: direct_bbl.tex
\providecommand{\href}[2]{#2}\begingroup\raggedright\endgroup

%% file: direct.bbl
\begin{thebibliography}{10}

\bibitem{d'Enterria:2012yj}
D.~d'Enterria and J.~Rojo, {\it {Quantitative constraints on the gluon
  distribution function in the proton from collider isolated-photon data}},
  Nucl. Phys. {\bf B860} (2012) 311; 
  arXiv:1202.1762.

\bibitem{Carminati:2012mm}
L.~Carminati, G.~Costa, D.~D'Enterria, I.~Koletsou, G.~Marchiori, {\em et~al.},
  {\it {Sensitivity of the LHC isolated-gamma+jet data to the parton
  distribution functions of the proton}},  Europhys. Lett. {\bf 101} (2013) 61002; 
  arXiv:1212.5511.

\bibitem{Aad:2012tfa}
ATLAS Collaboration, G.~Aad {\em et~al.}, {\it {Observation of a new
  particle in the search for the Standard Model Higgs boson with the ATLAS
  detector at the LHC}},  Phys. Lett. B {\bf 716} (2012) 1;
  arXiv:1207.7214.

\bibitem{Chatrchyan:2012ufa}
CMS Collaboration, S.~Chatrchyan {\em et~al.}, {\it {Observation of a new
  boson at a mass of 125 GeV with the CMS experiment at the LHC}}, 
  Phys. Lett. B {\bf 716} (2012) 30;
  arXiv:1207.7235.

\bibitem{Catani:2002ny}
S.~Catani, M.~Fontannaz, J.~Guillet, and E.~Pilon, {\it {Cross-section of
  isolated prompt photons in hadron-hadron collisions}}, JHEP {\bf 0205} (2002) 028;
  arXiv:hep-ph/0204023.

\bibitem{Aad:2010sp}
ATLAS Collaboration, G.~Aad {\em et~al.}, {\it {Measurement of the
  inclusive isolated prompt photon cross section in $pp$ collisions at
  $\sqrt{s}=7$ TeV with the ATLAS detector}},  Phys. Rev. D {\bf 83} (2011) 052005; 
  arXiv:1012.4389.

\bibitem{Aad:2011tw}
ATLAS Collaboration, G.~Aad {\em et~al.}, {\it {Measurement of the
  inclusive isolated prompt photon cross-section in $pp$ collisions at
  $\sqrt{s}=7$ TeV using 35 pb$^{-1}$ of ATLAS data}},  Phys. Lett. B {\bf 706}
  (2011) 150; 
  arXiv:1108.0253.

\bibitem{ATLAS:2012ar}
ATLAS Collaboration, G.~Aad {\em et~al.}, {\it {Measurement of the
  production cross section of an isolated photon associated with jets in
  proton-proton collisions at $\sqrt{s}=7$ TeV with the ATLAS detector}}, 
  Phys. Rev. D {\bf 85} (2012) 092014;
  arXiv:1203.3161.

\bibitem{Aad:2013gaa}
ATLAS Collaboration, G.~Aad {\em et~al.}, {\it {Dynamics of
  isolated-photon plus jet production in pp collisions at $\sqrt(s)=7$ TeV with
  the ATLAS detector}},  Nucl. Phys. {\bf B875} (2013) 483; 
  arXiv:1307.6795.

\bibitem{Aad:2013zba}
ATLAS Collaboration, G.~Aad {\em et~al.}, {\it {Measurement of the
  inclusive isolated prompt photons cross section in pp collisions at
  $\sqrt{s}=7$  TeV with the ATLAS detector using 4.6~fb$^{-1}$}},
  Phys. Rev. D {\bf 89} (2014), 052004;
  arXiv:1311.1440.

\bibitem{Khachatryan:2010fm}
CMS Collaboration, V.~Khachatryan {\em et~al.}, {\it {Measurement of the
  Isolated Prompt Photon Production Cross Section in $pp$ Collisions at
  $\sqrt{s} = 7$~TeV}}, Phys. Rev. Lett. {\bf 106} (2011) 082001; 
  arXiv:1012.0799.

\bibitem{Chatrchyan:2011ue}
CMS Collaboration, S.~Chatrchyan {\em et~al.}, {\it {Measurement of the
  Differential Cross Section for Isolated Prompt Photon Production in pp
  Collisions at 7 TeV}},  Phys. Rev. D {\bf 84} (2011) 052011; 
  arXiv:1108.2044.

\bibitem{Chatrchyan:2013mwa}
CMS Collaboration, S.~Chatrchyan {\em et~al.}, {\it {Measurement of the
  triple-differential cross section for photon+jets production in proton-proton
  collisions at $\sqrt{s}$=7 TeV}},  JHEP {\bf 1406} (2014) 009; 
  arXiv:1311.6141.

\bibitem{Buttar:2008jx}
C.~Buttar, J.~D'Hondt, M.~Kramer, G.~Salam, M.~Wobisch, {\em et~al.}, {\it
  {Standard Model Handles and Candles Working Group: Tools and Jets Summary
  Report}},  arXiv:0803.0678.

\bibitem{Kurihara:2002ne}
Y.~Kurihara {\em et~al.}, {\it {QCD event generators with next-to-leading order
  matrix elements and parton showers}},  Nucl. Phys. {\bf B654} (2003) 301; 
  arXiv:hep-ph/0212216.

\bibitem{Odaka:2007gu}
S.~Odaka and Y.~Kurihara, {\it Initial-state parton shower kinematics for NLO
  event generators},  Eur. Phys. J. C {\bf 51} (2007) 867; 
  arXiv:hep-ph/0702138.

\bibitem{Tsuno:2002ce}
S.~Tsuno {\em et~al.}, {\it {GR\@PPA\_4b: a four bottom quark production event
  generator for $pp/p\bar{p}$ collisions}},  Comput. Phys. Commun. {\bf 151} (2003) 216; 
  arXiv:hep-ph/0204222.

\bibitem{Tsuno:2006cu}
S.~Tsuno, T.~Kaneko, Y.~Kurihara, S.~Odaka, and K.~Kato, {\it {GR@PPA 2.7 event
  generator for $pp/p\bar{p}$ collisions}},  Comput. Phys. Commun. {\bf 175} (2006) 665; 
  arXiv:hep-ph/0602213.

\bibitem{Sjostrand:2006za}
T.~Sjostrand, S.~Mrenna, and P.~Z. Skands, {\it {PYTHIA 6.4 Physics and
  Manual}},  JHEP {\bf 05} (2006) 026; 
  arXiv:hep-ph/0603175.

\bibitem{Corcella:2000bw}
G.~Corcella {\em et~al.}, {\it {HERWIG 6: an event generator for hadron
  emission reactions with interfering gluons (including supersymmetric
  processes)}},  JHEP {\bf 01} (2001) 010; 
  arXiv:hep-ph/0011363.

\bibitem{Odaka:2011hc}
S.~Odaka and Y.~Kurihara, {\it {GR@PPA 2.8: Initial-state jet matching for weak
  boson production processes at hadron collisions}},  Comput. Phys. Commun. 
  {\bf 183} (2012) 1014; 
  arXiv:1107.4467.

\bibitem{Odaka:2012da}
S.~Odaka, {\it {GR@PPA 2.8.3 update}}, arXiv:1201.5702.

\bibitem{Odaka:2009qf}
S.~Odaka, {\it {Simulation of $Z$ boson $p_{T}$ spectrum at Tevatron by
  leading-order event generators}},  Mod. Phys. Lett. A {\bf 25} (2010) 3047; 
  arXiv:0907.5056.

\bibitem{Odaka:2012iz}
S.~Odaka, {\it {Simulation of $Z$-boson $p_{T}$ spectrum at LHC and Tevatron
  using GR@PPA}},  arXiv:1206.3398.

\bibitem{Odaka:2013fb}
S.~Odaka, {\it {Precise simulation of the initial-state QCD activity associated
  with $Z$-boson production in hadron collisions}},  Mod. Phys. Lett. A {\bf 28} (2013) 1350098; 
  arXiv:1301.5082.

\bibitem{Odaka:2012ry}
S.~Odaka and Y.~Kurihara, {\it {Consistent simulation of non-resonant diphoton
  production at hadron collisions with a custom-made parton shower}}, 
  Phys. Rev. D {\bf 85} (2012) 114022; 
  arXiv:1203.4038.

\bibitem{Odaka:2014ura}
S.~Odaka, N.~Watanabe, and Y.~Kurihara, {\it {ME-PS matching in the simulation
  of multi-jet production in hadron collisions using a subtraction method}},
  PTEP {\bf 2015} (2015) 053B04; 
  arXiv:1409.3334.

\bibitem{Hoeche:2009xc}
S.~Hoeche, S.~Schumann, and F.~Siegert, {\it {Hard photon production and
  matrix-element parton-shower merging}},  Phys. Rev. D {\bf 81} (2010) 034026;
  arXiv:0912.3501.

\bibitem{Aad:2012tba}
ATLAS Collaboration, G.~Aad {\em et~al.}, {\it {Measurement of
  isolated-photon pair production in $pp$ collisions at $\sqrt{s}=7$ TeV with
  the ATLAS detector}},  JHEP {\bf 1301} (2013) 086; 
  arXiv:1211.1913.

\bibitem{Catani:2001cc}
S.~Catani, F.~Krauss, R.~Kuhn, and B.~R. Webber, {\it {QCD Matrix Elements +
  Parton Showers}},  JHEP {\bf 11} (2001) 063; 
  arXiv:hep-ph/0109231.

\bibitem{Ishikawa:1993qr}
MINAMI-TATEYA group, T.~Ishikawa {\em et~al.}, {\it {GRACE
  manual: automatic generation of tree amplitudes in Standard Models: Version
  1.0}}, KEK-92-19 (1993).

\bibitem{Yuasa:1999rg}
F.~Yuasa {\em et~al.}, {\it {Automatic computation of cross sections in HEP:
  Status of GRACE system}},  Prog. Theor. Phys. Suppl. {\bf 138} (2000) 18; 
  arXiv:hep-ph/0007053.

\bibitem{Bourhis:1997yu}
L.~Bourhis, M.~Fontannaz, and J.~Guillet, {\it {Quarks and gluon fragmentation
  functions into photons}},  Eur. Phys. J. C {\bf 2} (1998) 529; 
  arXiv:hep-ph/9704447.

\bibitem{Kawabata:1985yt}
S.~Kawabata, {\it {A New Monte Carlo Event Generator for High-Energy Physics}},
   Comput. Phys. Commun. {\bf 41} (1986) 127.

\bibitem{Kawabata:1995th}
S.~Kawabata, {\it {A new version of the multidimensional integration and event
  generation package BASES/SPRING}},  Comp. Phys. Commun. {\bf 88} (1995)
  309.

\bibitem{Pumplin:2002vw}
J.~Pumplin {\em et~al.}, {\it {New generation of parton distributions with
  uncertainties from global QCD analysis}},  JHEP {\bf 07} (2002) 012; 
  arXiv:hep-ph/0201195.

\bibitem{Cacciari:2011ma}
M.~Cacciari, G.~P. Salam, and G.~Soyez, {\it {FastJet User Manual}}, 
  Eur. Phys. J. C {\bf 72} (2012) 1896; 
  arXiv:1111.6097.

\bibitem{Sherstnev:2007nd}
A.~Sherstnev and R.~S. Thorne, {\it {Parton Distributions for LO Generators}},
  Eur. Phys. J. C {\bf 55} (2008) 553; 
  arXiv:0711.2473.

\end{thebibliography}
